# A precisely regulating phase evolution strategy for highly efficient kesterite solar cells

Jiazheng Zhou†, Xiao Xu†, Huijue Wu, Jinlin Wang, Licheng Lou, Kang Yin, Yuancai Gong, Jiangjian Shi, Yanhong Luo, Dongmei Li*, Hao Xin*, Qingbo Meng*

J. Zhou, X. Xu, H. Wu, J. Wang, L. Lou, K. Yin, J. Shi, Y. Luo, D. Li, Q. Meng

Beijing National Laboratory for Condensed Matter Physics, Renewable Energy Laboratory

Institute of Physics, Chinese Academy of Sciences (CAS), Beijing 100190, P. R. China

E-mail: dmli@iphy.ac.cn, qbmeng@iphy.ac.cn

Q. Meng

Center of Materials Science and Optoelectronics Engineering, University of Chinese Academy of Sciences, Beijing 100049, P. R. China

Y. Gong, H. Xin

State Key Laboratory of Organic Electronics and Information Displays & Institute of Advanced Materials, Nanjing University of Posts & Telecommunications, Nanjing 210023 China

E-mail: iamhxin@njupt.edu.cn

J. Zhou, X. Xu, J. Wang, L. Lou, K. Yin, Y. Luo, D. Li, Q. Meng

School of Physical Sciences, University of Chinese Academy of Sciences, Beijing 100049, P. R. China

Y. Luo, D. Li, Q. Meng

Songshan Lake Materials Laboratory, Dongguan, Guangdong 523808, P. R. China

Abstract

Phase evolution during the selenization is crucial for high-quality kesterite $Cu_2ZnSn(S, Se)_4$ (CZTSSe) absorbers and efficient solar cells. Herein, we regulate kinetic process of phase evolution from $Cu^+$-$Sn^{4+}$-MOE (MOE: 2-methoxyethanol) system by precisely controlling positive chamber pressure. We found that, at the heating-up stage, Se vapor concentration is intentionally suppressed in low-temperature region, which effectively reduces collision probability between the CZTS and Se atoms, thus remarkably inhibiting formation of secondary phases on the surface and multiple-step phase evolution processes. This strategy enables the phase evolution to start at relatively higher temperature and thereby leading to high crystalline quality CZTSSe absorber with fewer defects, and corresponding CZTSSe solar cell can present 14.1% efficiency (total area), which is the highest result so far. This work provides important insights into selenization mechanism of CZTSSe absorbers and explores a new way of kinetic regulation strategy to simplify the phase evolution path to efficient CZTSSe solar cells.

As one of the emerging solar cells, kesterite $Cu_2ZnSn(S_xSe_{1-x})_4$ (CZTSSe) solar cell has presented 13% of the power convention efficiency (PCE) based on environmentally friendly solution systems, which have shown attractive application prospects[1]. However, large open-circuit voltage deficit ($V_{OC,\ def}$) still restricts the further improvement of the device performance, primarily due to complex phase composition and high deep-level defects in the CZTSSe absorber[2, 3, 4]. In fact, high-temperature selenization process is an indispensable step to complete the phase evolution of CZTSSe absorber, which is mainly affected by Se concentration[5, 6, 7, 8], reaction temperature and time[9, 10], even precursor film composition and chemical environments[11, 12], and so on. Besides, multi-element component CZTSSe has a narrow stable phase region, which usually undergoes complicated phase evolution pathways. Different phase evolution pathways imply dissimilar intermediate phases and different defect properties of the CZTSSe absorber[13, 14, 15, 16]. Therefore, precisely controlling CZTSSe phase evolution pathways to avoid detrimental intermediate phases is the key to high crystalline quality CZTSSe absorbers with low defects and pure kesterite phase.

Currently, some groups have already explored phase evolution process and crystal growth of the CZTSSe absorber during the selenization reaction, from different precursor compositions[14, 15, 16], local chemical environment[12], element doping[17, 18, 19, 20, 21], atmosphere regulation[22, 23, 24], to pressure controlling[25, 26, 27, 28], etc. These works contribute to understanding CZTSSe phase evolution and the reason for the $V_{OC,\ def}$ to some degree. They demonstrate that, at the early heating-up stage of the selenization reaction, gaseous Se easily reacted with the CZTS precursor film to give intermediate phases, *i.e.* binary phases ($Cu_xSe$, ZnSe, $SnSe_x$) and ternary phase ($Cu_2SnSe_3$, abbreviated as CTSe) with the temperature gradually increasing, depending on the composition of the precursor films derived from different preparation methods, e.g. vacuum method, nanocrystalline method or solution method[13, 23, 25, 29]. Besides, phase evolution pathways can directly affect the formation of intermediate phases (Fig. 1a). For example, for $Cu^+$-$Sn^{2+}$ involved precursors, its phase evolution process experienced successively from $Cu_2S$-ZnS-SnS to $Cu_2Se$-ZnSe-$SnSe_2$, then to CTSe-ZnSe, and finally to CZTSe (Path I)[15]. Saucedo et al introduced Ge(IV) to stabilize Cu-Sn alloy, which phase evolution process changed from metal stack to CTSe-ZnSe, then to CZTSe (Path II), and simultaneously avoided binary $Cu_xSe$ and $SnSe_x$ intermediate phases[14]. Xin et al used $Cu^+$-$Sn^{4+}$ as metal precursors in DMSO (dimethyl sulfoxide) system, which is supposed to directly change from CZTS to final CZTSe, excluding all intermediate phases (Path III)[15]. These works mentioned above all focus on phase evolution and significantly reduce $V_{OC,\ def}$, especially Xin's work, which achieves 13% efficient kesterite solar cell, instilling confidence in kesterite field[1]. Different from the CZTSe, intermediate phases mainly originated from Se participation in reactions at relatively lower temperatures, and reaction rates are so fast that could be completed within 2 min[30, 31]. Obviously, the time window for regulating intermediate phases is too narrow. On the other hand, most of high-efficiency CZTSSe devices were still based on a half-closed graphite box to complete selenization reaction, which could increase the difficulty of coordinating those parameters including temperature and Se vapor concentration[15, 20, 26]. Therefore, more effective regulation strategies are urgently required to avoid intermediate phases.

In this work, we regulate kinetic process of phase evolution by precisely tailoring a positive chamber pressure to reduce Se vapor concentration. The collision probability between the CZTS and gaseous Se molecule can be decreased in the low-temperature region while heating-up of the selenization reaction. Secondary phases resulting from decomposition on the surface and multi-step phase evolution paths are thus inhibited significantly. Additionally, this strategy enables the phase evolution to start at relatively higher temperature and thereby leading to high crystalline quality CZTSSe absorber with fewer defects. The bulk defects are reduced by around one order of magnitude. Finally, we achieved CZTSSe solar cell with 14.1% PCE (total area) and a certified 13.8% PCE (total area), which is the highest efficiency reported to date.

**Understanding the CZTSSe phase evolution process**

Selenization reaction of CZTS precursor films is the most important step among the fabrication processes of the CZTSSe solar cell, which covers phase evolution and crystal growth processes. Particularly, the phase evolution process is crucial to the CZTSSe crystal quality and defect formations. In this work, we prepared CZTS precursor films based on $Cu^{+}$-$Sn^{4+}$-MOE precursor solution (MOE: 2-methoxyethanol) in air ($CZTS_{air}$), which is supposed to be more suitable for mass production. When this $CZTS_{air}$ film was selenized under ambient atmospheric pressure, the occurrence of hetero-nucleation and intermediate phases was traced by interrupting at different time points. The heating-up curve was set as follows: elevating temperature from room temperature to 350°C in 60 s, then holding at 350°C for 300 s, finally continuously increasing to 535°C at a heating-up rate of 0.63°C/s. In this heating-up process, the $Cu_2ZnSn(S_xSe_{1-x})_4$ (CZTSSe) was obtained with a Raman peak at 175 $cm^{-1}$, and other peaks varied from 215 to 196 $cm^{-1}$ as $x$ changing from 0.5 to 0 while the temperature was increasing (Fig. 1b)[32] (Supplementary Note 1 about Raman peak fitting). In the meantime, however, slight elemental Se (trigonal Se: 233 $cm^{-1}$, a-Se: 250 $cm^{-1}$) was left on the sample surface when the reaction was prematurely terminated (Supplementary Fig. S1)[33]. Typically, Raman peak at 180 $cm^{-1}$ assigned to the CTSe phase, was also found when the selenization temperature reached 450°C [34]. The peak area increases as the temperature rising from 450 to 510°C but disappears when further increasing the temperature to 535°C. It is thus suggested that, in this case, the phase evolution may undergo Path I or Path II. The formation of the CTSe is mainly related to the precursor film, that is, oxygen-metal bonds are easily formed by thermally treating the CZTS precursor film in the air, especially for alcohol-based (R-OH) solvent systems [35, 36]. This can be further confirmed by the O1s peak in Ar-etched XPS (~ 100 nm etching depth) by comparison of MOE-air-precursor and DMSO-$N_2$-precursor (Supplementary Fig. S2). The precursor film is generally considered to be poor-crystallized Cu-Zn-Sn-S along with a small amount of crystalline CZTS, based on XRD pattern and Raman spectra (Supplementary Fig. S3). In terms of thermodynamics, this precursor film could easily react with Se vapor during the selenization reaction, leading to some CZTS decomposition accompanied with the appearance of binary or ternary phases, which may make the phase evolution path complicated[29], even shifting from Path III to Path I or II.

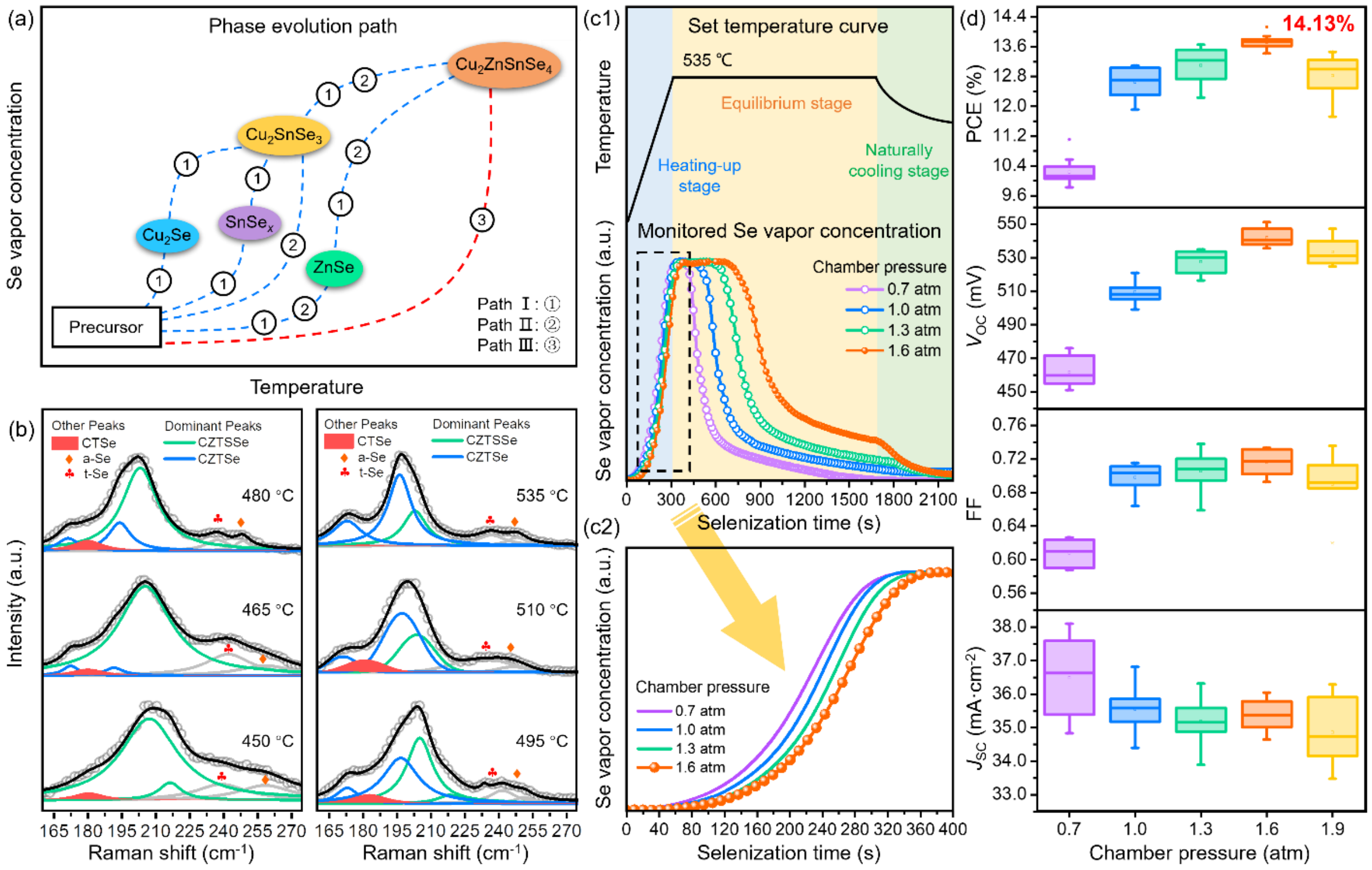

**Fig. 1. The method to regulate kinetic process of phase evolution and the resulted solar cell performance.** (a) schematic diagram of phase evolution path (normalized Se vapor concentration *vs*. temperature); (b) Raman spectra (532 nm excitation wavelength) of samples obtained at 450, 465, 480, 495, 510, and 535°C, with relatively slow heating-up rate of 0.63°C/s under ambient pressure; (c1) variation trend of the temperature, Se vapor concentration *vs*. selenization time under various chamber pressures derived from real-time monitoring results; (c2) enlarged view of the heating-up stage in Fig. c1; (d) statistical results of photovoltaic parameters (all based on total area with anti-reflection coating): PCE, $V_{OC}$, fill factor (FF), and short-circuit current density ($J_{SC}$).

## *In-situ* monitoring and tailoring kinetic phase evolution

Tailoring the chamber pressure is an effective way to obtain different Se vapor concentration. According to the semi-empirical relationship proposed by Gilliland, assuming that binary mixed gases ($N_2$/Se(g)) are ideal gases, the diffusion coefficient of gas molecules is proportional to $p^{-1}$ (pressure) and $T^{2/3}$ (temperature), based on molecular dynamics theory[37]. Theoretically, positive chamber pressure can decrease the gaseous Se diffusion rate to a certain extent, and vice versa. In order to verify experimentally, the Se vapor concentration in the selenization reaction was *in-situ* monitored by our lab-made selenization furnace, detailed measurement process is given in Supplementary Note 2 and Fig. S4. Variation tendencies of Se concentration with the reaction time under different positive pressures are presented in Fig. 1c1, and all the curves are divided into three parts corresponding to heating-up, equilibrium and naturally cooling stages in the whole selenization process. With the chamber pressure increasing, the Se vapor concentration rising rate is reduced at the heating-up stage but the time to reach stable Se vapor value is distinctly prolonged, well consistent with the above speculation (Fig. 1c2). Therefore, regulating the chamber pressure mainly affects the early stage in selenization reaction.

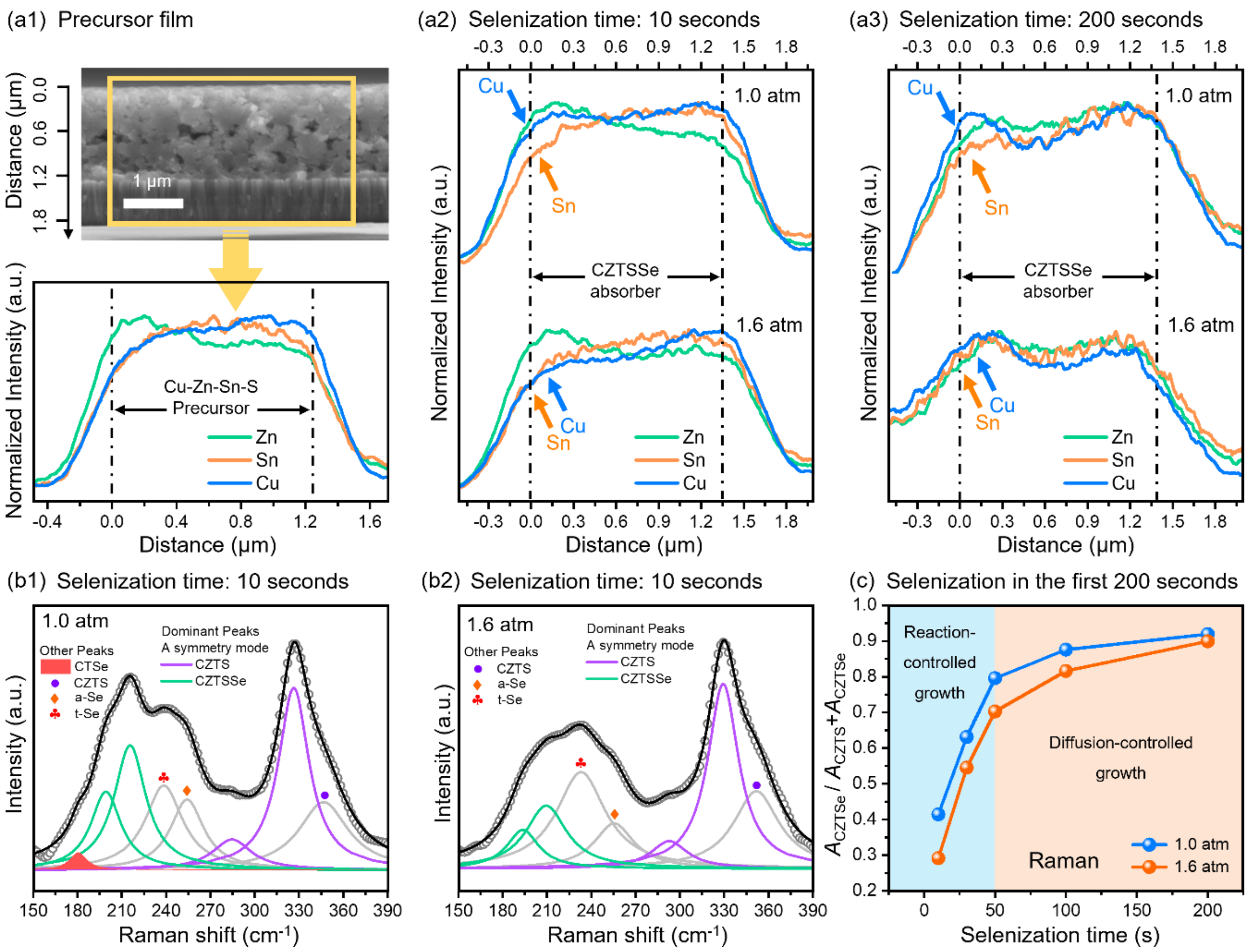

**Fig. 2. Influence of chamber pressure on kinetic process of phase evolution.** The samples obtained under various conditions are denoted as *p*-535-*t* (*p*: pressure (atm); *t*: selenization time (s)). (a) EDX element mappings (normalize by the maximum value) of Cu-Zn-Sn-S precursor (a1), 1.0-535-10 (a2), and 1.6-535-10 (a3); (b) Raman spectra (532 nm excitation wavelength) of 1.0-535-10 (b1), 1.6-535-10 (b2); (c) $R_{\text{Raman}}$ of *p*-535-*t* obtained by Raman spectra *vs*. selenization time in the early 200 s, *p*=

1.0 and 1.6, $t$= 10, 30, 50, 100, and 200. $R_{Raman}$ is the peak area ratio of CZTSe phase to the sum of CZTS and CZTSSe phases, $R_{Raman}= A_{CZTSe}/(A_{CZTS}+A_{CZTSe})$, $A$: Raman peak area of CZTS phase (290-350 $cm^{-1}$) or CZTSSe phase (160-220 $cm^{-1}$).

Based on CZTSSe absorbers under different chamber pressure, we further fabricated CZTSSe solar cells with a configuration of Mo/CZTSSe/CdS/ZnO/ITO/Ni-Al/$MgF_2$. According to statistical results of photovoltaic parameters (all based on total area) in Fig. 1d, with the chamber pressure increasing, both FF and $V_{OC}$ are remarkably improved whereas the $J_{SC}$ is slightly decreased mainly due to different band gap by the Se/(S+Se) ratio (XRF data in Supplementary Fig. S5), and the average PCE reaches to the highest value of 13.72% under 1.6 atm. This variation tendency reflects the positive chamber pressure indeed could bring better cell performance of CZTSSe solar cells.

**Characteristics of phase evolution and crystal growth**

Influence of different positive pressures on the early phase evolution of CZTSSe has been explored firstly. Under different chamber pressures, the $CZTS_{air}$ film was selenized by directly raising the temperature to 535°C in 60 s, then holding at 535°C for 1300 s, finally naturally cooling to room temperature. Based on the above discussion, the rapid heating-up rate (60 s to 535°C) was set to avoid intermediate phases (Fig. 1b), here, we focus on the reaction stage at 535°C. EDX element mappings show that, for the $CZTS_{air}$ precursor film, Zn is slightly enriched on the surface while the distributions of Cu and Sn are basically uniform (Supplementary Fig. S6, Fig. 2a1). However, for samples selenized under ambient pressure (Fig. 2a2-2a3), the diffusion rate of Cu from film bottom to surface is higher than that of Sn, thus leading to a Cu-Zn-rich and Sn-depleted surface. It is supposed that, under ambient pressure, the reaction between CZTS and Se vapor at the film surface, may cause decomposition of CZTSSe, and the $Cu_xSe$ simultaneously forms in the initial heating-up stage. Therefore, the $Cu_xSe$ formation process will promote the Cu diffusion to the surface. These two processes mutually reinforce each other, resulting in the above difference. In this way, the phase evolution in the entire film shows a characteristic of one-step transition but with a small amount of intermediate phases on the surface, that is, most Cu-Zn-Sn-S in bulk precursor reacts with Se to directly transform CZTSSe phase, but a little Cu-Zn-Sn-S on the surface decomposes and transforms from $Cu_xSe$ to CTSe, and finally CZTSSe phase. For positive samples (Fig. 2a2-2a3), due to much lower Se vapor concentration in the early stage, the thermodynamic condition for forming $Cu_xSe$ cannot be reached, therefore, the entire film is directly transformed from Cu-Zn-Sn-S to CZTSe with the formation of secondary phase can be well avoided.

Raman spectra further confirm this conclusion (Fig. 2b1 and 2b2), that is, the CTSe peak is detected in 1.0-535-10s samples, but cannot be detected in positive pressure samples (*i.e.,* 1.6 atm) (Fig. 1b, 2a2, 2b2, and Supplementary Fig. S7). Obviously, positive chamber pressures in the selenization process can efficiently retard the formation of those thermodynamic priority intermediate phases (*i.e.* CTSe), also confirmed by XRD patterns (Supplementary Fig. S8). The effect of positive pressures on the early phase evolution process (within early 200 s), mainly lies in selenization reaction degree, which is evaluated by a semi-quantitative method: the $R_{Raman}$ value ($R_{Raman}= A_{CZTSe}/(A_{CZTS}+A_{CZTSe})$) (Fig. 2c). The less the $R_{Raman}$ value is, the slower conversion from amorphous CZTS to crystallized CZTSSe is, and thus the slower reaction rate is. It can be seen from Fig. 2c that, in early 200 s' reaction stage, the $R_{Raman}$ under positive chamber pressures is lower than that under ambient pressure, but can reach similar value at 200 s. It is thus proposed that the starting time of the phase evolution has been delayed under positive pressures, and the difference in phase evolution is mainly reflected in the very earlier reaction stage. It is worth noting that the $R_{Raman}$ versus selenization time exhibits characteristic double exponential decay, also suggesting a rapid decay in the early 50 s (blue area), followed by a slow decay process (orange area). This is consistent with the crystal growth theory proposed in early work, namely, the crystal growth is involved by two stages, one is a reaction-controlled growth stage with obvious crystal orientation growth, the other is a diffusion-controlled growth stage with round grains on the surface[29].

In our work, the reaction-controlled growth stage only lasts about 50 s, whereas subsequent diffusion-controlled growth stage will last till ending. In fact, the most critical phase evolution process is determined by reaction-controlled growth stage, therefore, how to precisely regulate this reaction-controlled growth stage becomes crucial. Based on the above discussion, we reach a conclusion that increasing chamber pressure can inhibit the earlier reaction between Se vapor and the precursor, and the early phase evolution stage is delayed around 30 to 50 s, therefore, intermediate phases and Cu, Sn heterogeneous diffusions will be effectively avoided, and the kinetic regulation window of CZTSSe phase can also be widened. This strategy is beneficial for regulating the overall phase evolution pathway.

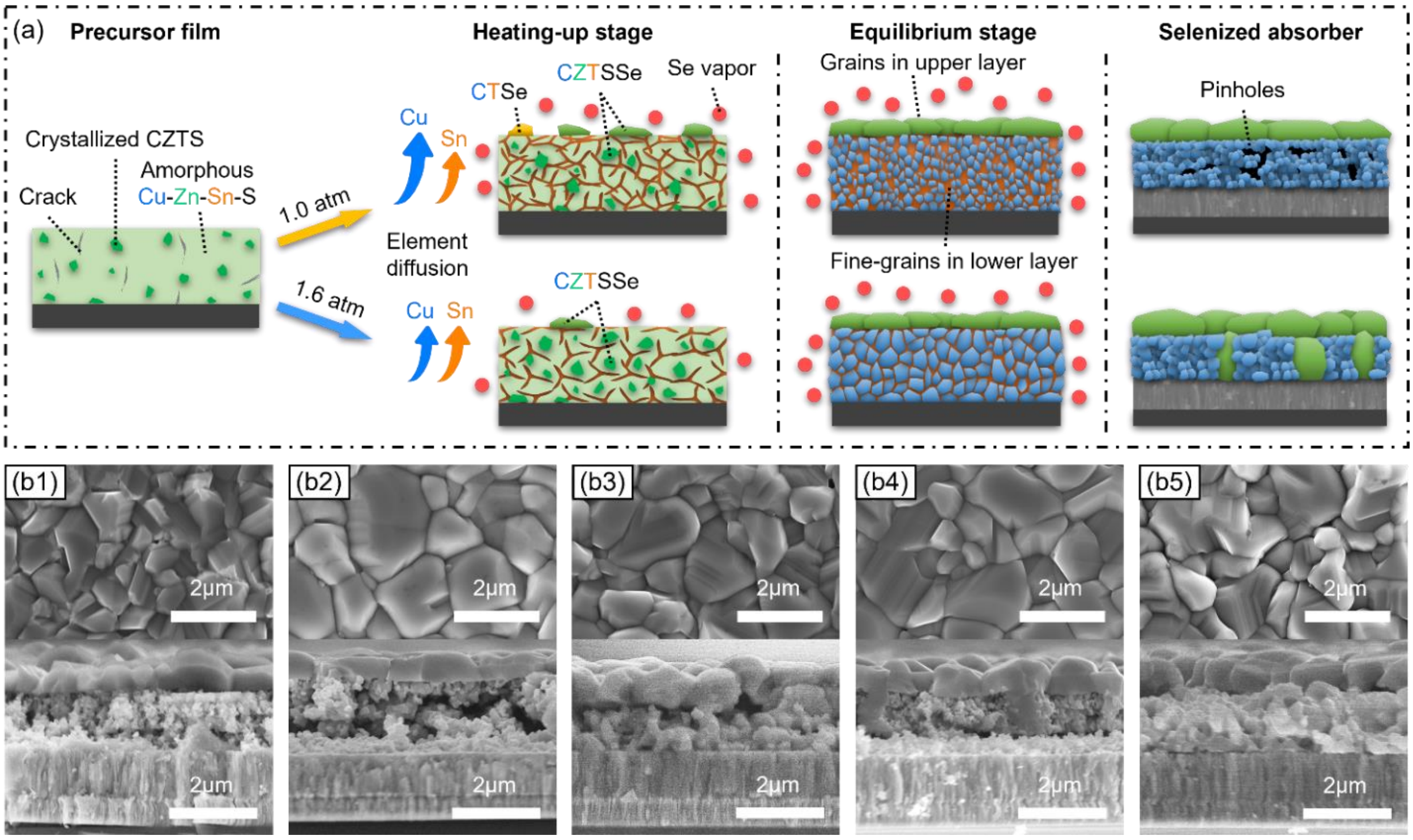

**Fig. 3. Influence of chamber pressure on CZTSSe morphology.** (a) Schematic diagram for the selenization process; (b) Top-view and cross-section SEM images of final CZTSSe absorbers under 0.7 atm (b1), 1.0 atm (b2), 1.3 atm (b3), 1.6 atm (b4), and 1.9 atm (b5) chamber pressure.

A schematic diagram is further suggested about the crystal growth mechanism in selenization process by regulating positive chamber pressures (Fig. 3a). For element diffusion and phase evolution on the surface, it is consistent with the above discussion. For crystal growth, however, sufficient Se vapor uniformly penetrates into the precursor film with loose, porous surface, then crystallization occurs preferentially in the upper film with island-like grains. Small grains are finally mixed and ripen to afford large grains. In fact, nucleation and crystallization occur simultaneously on the surface and inside precursor film, however, fine grains are formed in the film bottom, which may bring much more grain boundaries and bulk defects even unsatisfied device efficiency[38]. When the chamber pressure is increased, some fine grains in the bottom can fuse to larger grains with less pores (Fig. 3b1-b5 and Supplementary Fig. S9). When 1.6 atm of the chamber pressure is adopted, relatively larger grains in the film bottom and better contact with the Mo substrate are obtained (Fig. 3b4). This dense and less-pinholes morphology in the film bottom can reduce grain boundaries and resulting defects, improving the back interface property.

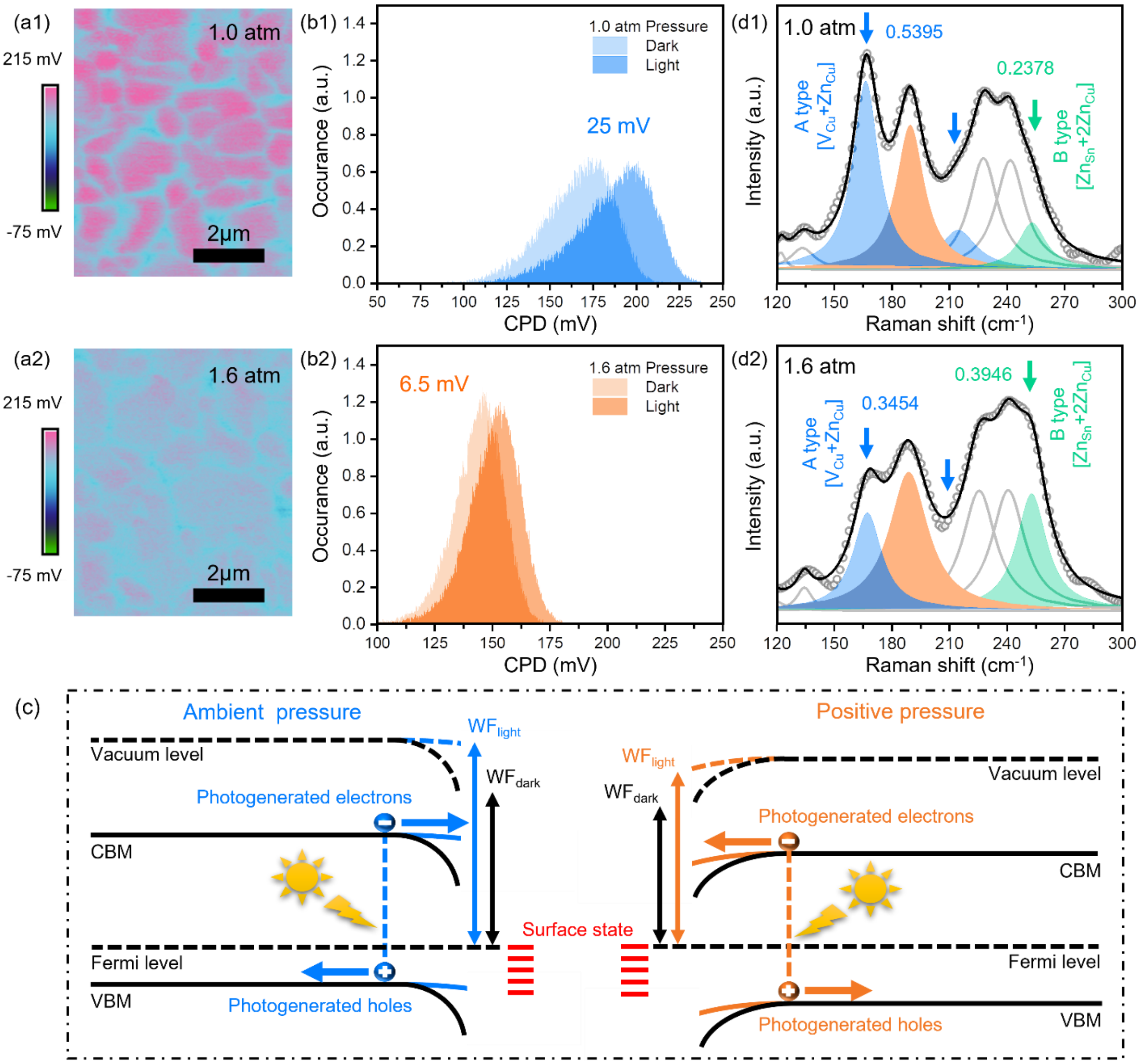

**Fig. 4. Surface state and defect properties.** (a) KPFM images of 1.0-CZTSSe (a1) and 1.6-CZTSSe (a2); (b) contact potential difference (CPD) distribution of 1.0-CZTSSe (b1) and 1.6-CZTSSe (b2) in dark and under illumination; (c) Schematic energy band diagram for CZTSSe surface under ambient and positive pressure (CBM, conduction band minimum; VBM, valence band maximum; $WF_{dark}$, work function of the CZTSSe in dark; $WF_{light}$, work function of the CZTSSe under illumination;). The solid lines and dotted lines represent dark and light conditions, respectively; (d) Raman spectra (325 nm excitation wavelength) of final CZTSSe absorbers: 1.0-CZTSSe (d1) and 1.6-CZTSSe (d2).

**Surface defect properties and device performances**

Effect of positive chamber pressure on surface states and defect properties of final CZTSSe absorbers, 1.0-CZTSSe and 1.6-CZTSSe (chamber pressure: 1.0 and 1.6 atm), are explored. The potential profiling demonstrates that positive chamber pressure results in the downshift in Fermi level of the CZTSSe film and relatively higher shallow acceptor carriers, as ~50 mV lower potential value of the 1.6-CZTSSe absorber than that of 1.0-CZTSSe absorber (Fig. 4b-c, related AFM images in Supplementary Fig. S10). On the other hand, a positive chamber pressure can also bring much more uniform potential of grain surface and grain boundary, indicating both impurity phases and defects at grain boundary are significantly reduced. Moreover, based on previous works, relatively lower difference between relative potential values in dark and under illumination means less surface states on the CZTSSe surface (Fig. 4c-d, detailed discussion in Supplementary Note 4)[39]. Obviously, under the positive chamber pressure, the crystallinity of the CZTSSe absorber has been significantly improved.

Surface defect property of as-obtained CZTSSe is also investigated by Raman spectra (325 nm excitation wavelength), which usually can detect ~ 30 nm depth from the film surface. Peak area ratios denoted as $A_{168}/(A_{168}+A_{190})$ and $A_{253}/(A_{253}+A_{190})$, represent point-defect clusters of A type [$V_{Cu}$+$Zn_{Cu}$] and B type [$Zn_{Sn}$+2$Zn_{Cu}$], respectively, where characteristic CZTSe phase (190 $cm^{-1}$) and Cu, Zn and Sn related defects (168, 230 and 253 $cm^{-1}$) are used (Fig. 4a1-a2)[40]. The CZTSSe absorber derived from positive chamber pressures, has relatively higher $V_{Cu}$ concentration due to lower $A_{168}/(A_{168}+A_{190})$ (1.0-CZTSSe: 0.5395, 1.6-CZTSSe: 0.3454, and detailed discussion in Supplementary Note 3). Moreover, deep-level defect $Sn_{Zn}$ concentration in the 1.6-CZTSSe is also decreased due to obvious increase of the $A_{253}/(A_{253}+A_{190})$ (1.0-CZTSSe: 0.2378, 1.6-CZTSSe: 0.3946). Therefore, less deep level defect is obtained for the 1.6-CZTSSe absorber, thus beneficial for reducing carrier recombination and $V_{OC}$ deficit to further improve device efficiency of kesterite solar cell.

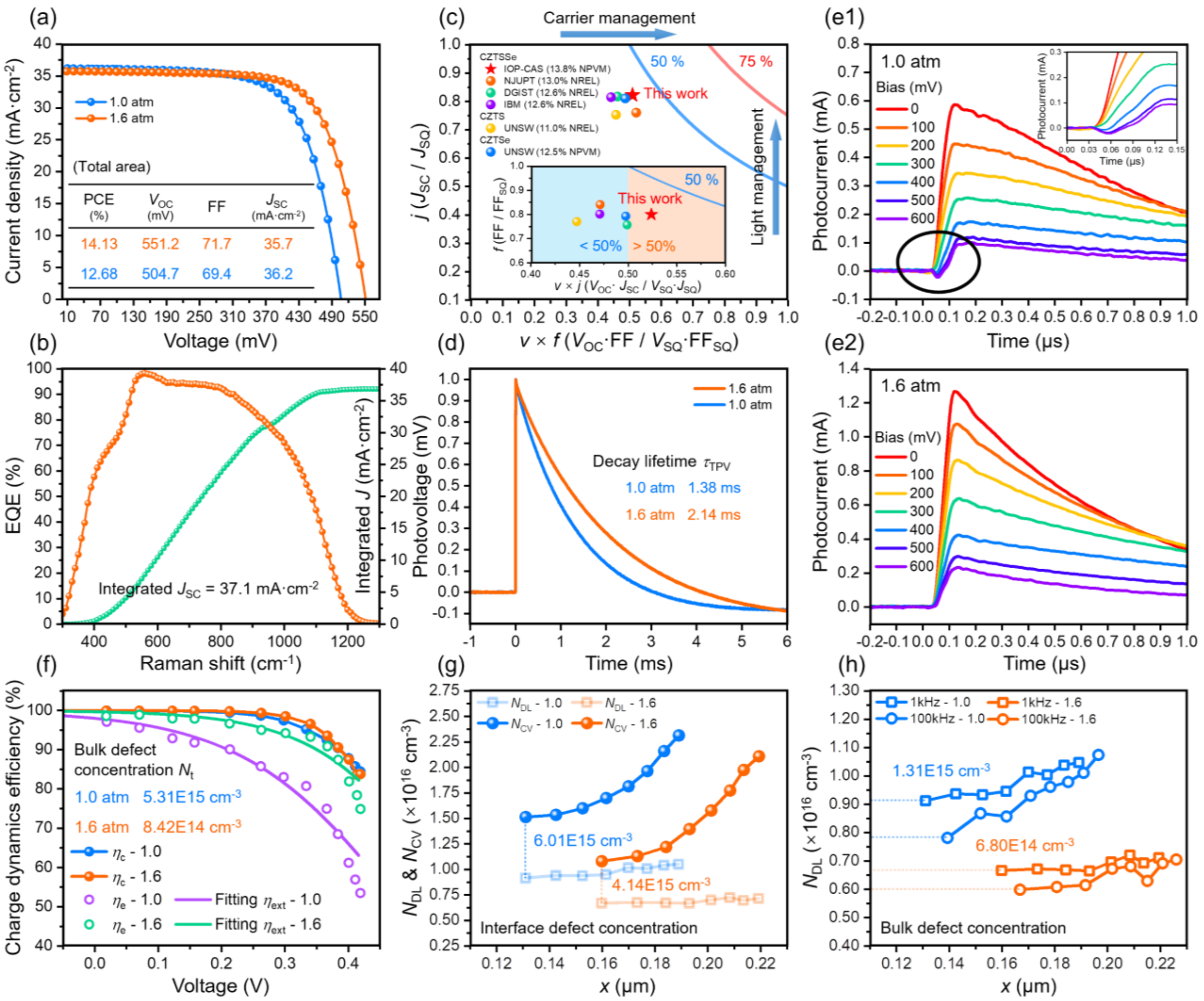

**Fig. 5. Performances of photovoltaic devices.** (a) the *J*-*V* curves of *p*-device (*p*= 1.0 and 1.6 atm); (b) EQE curve and integrated $J_{SC}$ of the best device; (c) fraction of Shockley-Queisser detailed-balance limit for $V_{OC}$, $J_{SC}$, and FF achieved by record kesterite cells and our result. The parameter $v \times f$ represents carrier management and *j* represents light management. The inset image also reflects S-Q limit information, but with $v \times j$ as horizontal axis and *f* as vertical axis; (d) transient photovoltage (TPV) spectra of *p*-device (*p*= 1.0 and 1.6 atm); (e) transient photocurrent (TPC) spectra of 1.0-device (e1) and 1.6 device (e2) under various forward biases; (f) modulated-TPV/TPC spectra of *p*-device (*p*= 1.0 and 1.6 atm); (g) DLCP ($N_{DL}$) and *C*-*V* ($N_{CV}$) of *p*-device (*p*= 1.0 and 1.6 atm) under 1 kHz; (h) DLCP ($N_{DL}$) of *p*-device (*p*= 1.0 and 1.6 atm) under 1 and 100 kHz.

The device based on 1.6-CZTSSe absorber presents 35.74 mA·cm$^{-2}$ of $J_{SC}$, 551.20 mV of $V_{OC}$, 71.73 % of FF, yielding the best 14.13 % PCE (Fig. 5a) and a certified 13.8 % PCE (Supplementary Fig. S11). Integrated $J_{SC}$ from EQE spectra is 37.1 mA·cm$^{-2}$ (without shield of grid lines), in good agreement with the active-area $J_{SC}$ (37.1 mA·cm$^{-2}$) (Fig. 5b). Its band gap energy ($E_g$) is 1.097 eV, estimated from the first derivative of the EQE curve in long wavelength (Supplementary Fig. S12). The $V_{OC}$ deficit can be reduced to 0.3042 eV ($V_{OC}^{SQ}$ -$V_{OC}$). This result is much better than the record kesterite cells (Fig. 5c), especially better carrier transportation property (less recombination and better transportation)[12, 22, 41, 42, 43]. As it is usually hard to simultaneously obtain high $V_{OC}$ and $J_{SC}$ in kesterite solar cells, the parameter $v \times j$ can be used to reflect the potential for absorber materials. For our devices, the parameter $v \times j$ is above 50% of $V_{SQ} \times J_{SQ}$ for the first time, however, for the other devices, this parameter is usually below 50%, suggesting better prospect for this kinetic regulation strategy.

Comparison of transient photovoltage (TPV) spectra reveals that, carrier recombination loss in the 1.6-CZTSSe based device has been reduced and longer $\tau_{TPV}$ is obtained (1.38 *vs*. 2.14 ms, under 0 V bias) (Fig. 5d). For transient photocurrent (TPC) spectra (Fig. 5e), however, the 1.0-device is an exception case since negative photocurrent signals appear for about 0.03 μs before rapid rising when > 400 mV forward bias is applied, not for 1.6-device. This further confirms much better back interface contact property of 1.6-device than that of 1.0-device[20]. Modulated-TPV/TPC spectra also demonstrate that the charge extraction efficiency ($\eta_E$) has been significantly improved as chamber pressure varying from the ambient to positive, consisting with the defects of bulk CZTSSe absorber is reduced and the recombination is suppressed as well[44, 45, 46]. The bulk defect state densities ($N_t$) are estimated to be $5.31 \times 10^{15}$ and $8.42 \times 10^{14}$ cm$^3$ for 1.0-device and 1.6-device, respectively (Fig. 5f, detailed discussion in Supplementary Note 5), basically consistent with their surface properties (Fig. 4).

Fig. 5g-h present interface defects and bulk defects derived from carrier concentrations by DLCP ($N_{DL}$) and CV ($N_{CV}$) under different frequency[47]. Interface defect concentrations of 1.0-device and 1.6-device are $6.01 \times 10^{15}$ and $4.14 \times 10^{15}$ cm$^3$, respectively, which are derived from the $N_{DL}$ and $N_{CV}$ in low frequency (1 kHz). Bulk defect concentrations of 1.0-device and 1.6-device are $1.31 \times 10^{15}$ and $6.80 \times 10^{14}$ cm$^3$, respectively, which can be described by the difference between $N_{DL}$ in low frequency (1 kHz) and $N_{DL}$ in high frequency (100 kHz). Obviously, positive pressures indeed help to reduce bulk defect concentration by about one order of magnitude as well as interfacial defects, finally leading to high-quality CZTSSe absorbers and high-performance photovoltaic devices.

**Conclusions**

We have precisely regulated the phase evolution process by collaboratively tailoring early selenization reaction rate in the half-closed graphite box. By increasing positive chamber pressure, Se vapor concentration has been intentionally decreased, which can also reduce the collision probability between CZTS precursor film and gaseous Se molecule in low-temperature region during the heating-up stage of selenization. Starting from regulating the kinetic process of the reaction, the decomposed secondary phases ($Cu_xSe$) on the surface and resulting multi-step phase evolution paths are thus inhibited significantly. This strategy enables the phase evolution to start at relatively higher temperature and thereby obtaining high crystalline quality CZTSSe absorber with fewer defects on the surface. The surface states are reduced, and the bulk defect density is reduced for around one order of magnitude (from $5.31\times10^{15}$ to $8.42\times10^{14}$ cm$^{3}$). Finally, we achieved CZTSSe solar cell with a self-measured PCE of 14.1% (total area) and a certified PCE of 13.8% (total area), which is the highest efficiency reported to date. This work provides a new way of kinetic regulation strategy for further understanding and regulating the phase evolution process of CZTSSe, especially optimizing the phase evolution path to achieve high-efficiency CZTSSe solar cells.

**Acknowledgements**

This work was financially supported by Natural Science Foundation of China (Nos. U2002216, 51972332, 52172261).

**Author contributions**

Jiazheng Zhou and Xiao Xu contributed equally to this work. The manuscript was written through contribution of all authors. And all authors have approval to the final version of the manuscript. Qingbo Meng, Hao Xin, Dongmei Li: Supervision, Discussion, Writing – review & editing. Jiazheng Zhou and Xiao Xu: Experiments, Characterization, Writing – original draft. Huijue Wu: Discussion, Se vapor concentration and IPCE measurement. Jinlin Wang, Licheng Lou, Kang Yin, Yuancai Gong: Data analyses and discussion. JiangJian Shi: Discussion, m-TPC/TPV analyses. Yanhong Luo: Data analyses.

**Competing interests**

There is no interest conflict.